\documentclass[aps,twocolumn,epsfig,graphics,showpacs,floatfix,superscriptaddress]{revtex4}
\usepackage{amsmath,amsfonts,amssymb,graphics,graphicx,epsfig,color,times,bbm}
\usepackage[latin1]{inputenc}

\newcommand{\ket}[1]{\left| #1 \right.\rangle}

\renewcommand{\H}[1]{\hat{H}_{\rm #1}}

\newcommand{\ad}{\hat{b}^\dagger}
\renewcommand{\a}{\hat{b}}
\newcommand{\nb}{\hat{n}}

\newcommand{\nf}{\hat{m}}

\DeclareMathOperator{\Tr}{Tr}

\begin{document}

\title{The one-dimensional Bose-Fermi-Hubbard model in the limit of fast fermions}

\author{Alexander Mering and Michael Fleischhauer}
\affiliation{Department of Physics and Research Center OPTIMAS, University of Kaiserslautern, 67663 Kaiserslautern, Germany}


\begin{abstract}
We discuss the ground-state phase diagram of the one-dimensional Bose-Fermi-Hubbard model (BFHM) 
in the limit of fast fermions based on an effective boson model. We give a detailed derivation of the effective model with long-range RKKY-type interactions, discuss its range of validity
and provide a deeper insight into its implications. In particular we show that integrating out the fast fermion degrees of freedom in a naive way
results in an ill-behaved effective Hamiltonian and a proper renormalization is required.  Based on the effective Hamiltonian, the phase diagram in the thermodynamic limit is constructed by analytic means and is compared to numerical results obtained by density matrix renormalization group (DMRG) techniques for the full BFHM. The most prominent 
feature of the phase diagram, the existence of a phase separation between Mott insulator (MI) and charge density 
wave (CDW) is discussed in depth including boundary effects. 
\end{abstract}


\date{\today}
\maketitle


\section{Introduction}

The advancement of quantum-optical tools during the last decades has made ultracold atoms in optical lattices an important and versatile experimental testing ground for quantum many-body phenomena of condensed matter physics. Recently systems with long-range interactions have gained substantial interest as the competition between local- and long-range interactions as well as the free motion of the particles can give rise to interesting many-body states including
 peculiar forms of quantum matter such as a supersolid, predicted 50 years ago
\cite{Thouless1969,Andreev1969,Leggett1970}, where superfluidity coexists with a non-vanishing structure factor.
 As shown in different theoretical works, supersolids can form in bosonic systems in the presence of non-local interactions \cite{vanOtterlo1995,Batrouni2000,Sengupta2005,Capogrosso-Sansone2010,Mishra2009}. The latter can be either intrinsic or they are
mediated through the interaction with a second species. The latter is the case for a mixture of bosons with spin polarized fermions, described by the Bose-Fermi-Hubbard (BFHM) model, in the limit of fast fermions.  For mixtures of bosons and fermions, H\'ebert \emph{et al.} showed by numerical means, that a supersolid of the bosons is present for half filling of fermions and if the bosons are doped away from half filling \cite{Hebert2008}.
Beside a supersolid, a multitude of other phases in mixed systems such as phase separation \cite{Batrouni2000,Sengupta2005,Mathey2007,Hebert2007,Orth2009,Titvinidze2008,Pollet2004} or incompressible charge-density wave (CDW) phases \cite{Mathey2007,Titvinidze2008,Pollet2004,Altman2003,Pollet2006,Mering2010} have been predicted. . 

Here we extend our previous work of \cite{Mering2010} and provide an analytic theory to understand the physics of the bosonic subsystem in the BFHM for fast fermions at half filling. The limit of fast fermions is of natural interest, since in most experimental realizations the fermionic atoms have a smaller effective mass, respectively a larger tunneling amplitude than the bosonic ones \cite{Guenther2006,Ospelkaus2006}. Following ideas in  \cite{Buechler2003} and adiabatically eliminating the fermions similar to the approach in \cite{Lutchyn2008} we derive an effective  bosonic Hamiltonian for $J_F\to\infty$, resulting in RKKY-type long-range couplings between bosons.
After explaining the nature of this mediated interaction, we discuss the bosonic phase diagram and discuss effects from spatial boundaries. 
All results are accompanied by numerical studies using DMRG for the full model.

The framework of our approach is set by the BFHM, describing a mixture of ultracold bosons and fermions in an optical lattice \cite{Albus2003}:
\begin{eqnarray}\label{eq:BFHM}
\hat H&=&-J_B\sum_j\left(\ad_j\a_{j+1}+\ad_{j+1}\a_{j}\right)+\frac{U}{2}\sum_j\nb_j\left(\nb_j-1\right)\nonumber\\
	&&-J_F\sum_j\left(\hat c^\dagger_j\hat c_{j+1}+ \hat c^\dagger_{j+1}\hat c_{j}\right)+V\sum_j\nb_j\nf_j,
\end{eqnarray}
Here, $\hat b^\dagger, \hat b$ ($\hat c^\dagger,\hat c$) are bosonic (fermionic) creation and annihilation operators and $\nb$ ($\nf$) the corresponding number operators. The bosonic (fermionic) 
hopping amplitude is given by $J_B$ ($J_F$), and $U$ ($V$) accounts for the intra- (inter-) species interaction energy. In the following
we restrict ourselves to the limit of large fermionic hopping, i.e. we assume $J_F\gg U,|V|,J_B$ and the energy scale is set by $U=1$. \\

\section{Mean-field approximation of fermions}\label{chap:Friedel}

\subsection{infinite system}

A first, intuitive ansatz to understand the physics in the regime of ultrafast fermions is to assume of a full decoupling of the fermions from the bosons. 
This assumption leads to a homogeneous fermion distribution $\left\langle \nf_j\right\rangle = \rho_F$ and the effective potential arising from the the interaction part
\begin{equation}
 V \sum_j \nb_j \nf_j \to V \rho_F\sum_j \nb_j
\end{equation}
simply gives a shift of the bosonic chemical potential as $\mu_B\mapsto \mu_B-V\rho_F$. In this limit the bosonic sub-system maps to
the Bose-Hubbard model (BHM) with a modified chemical potential.

\begin{figure}[h]%
\includegraphics*[width=\linewidth]{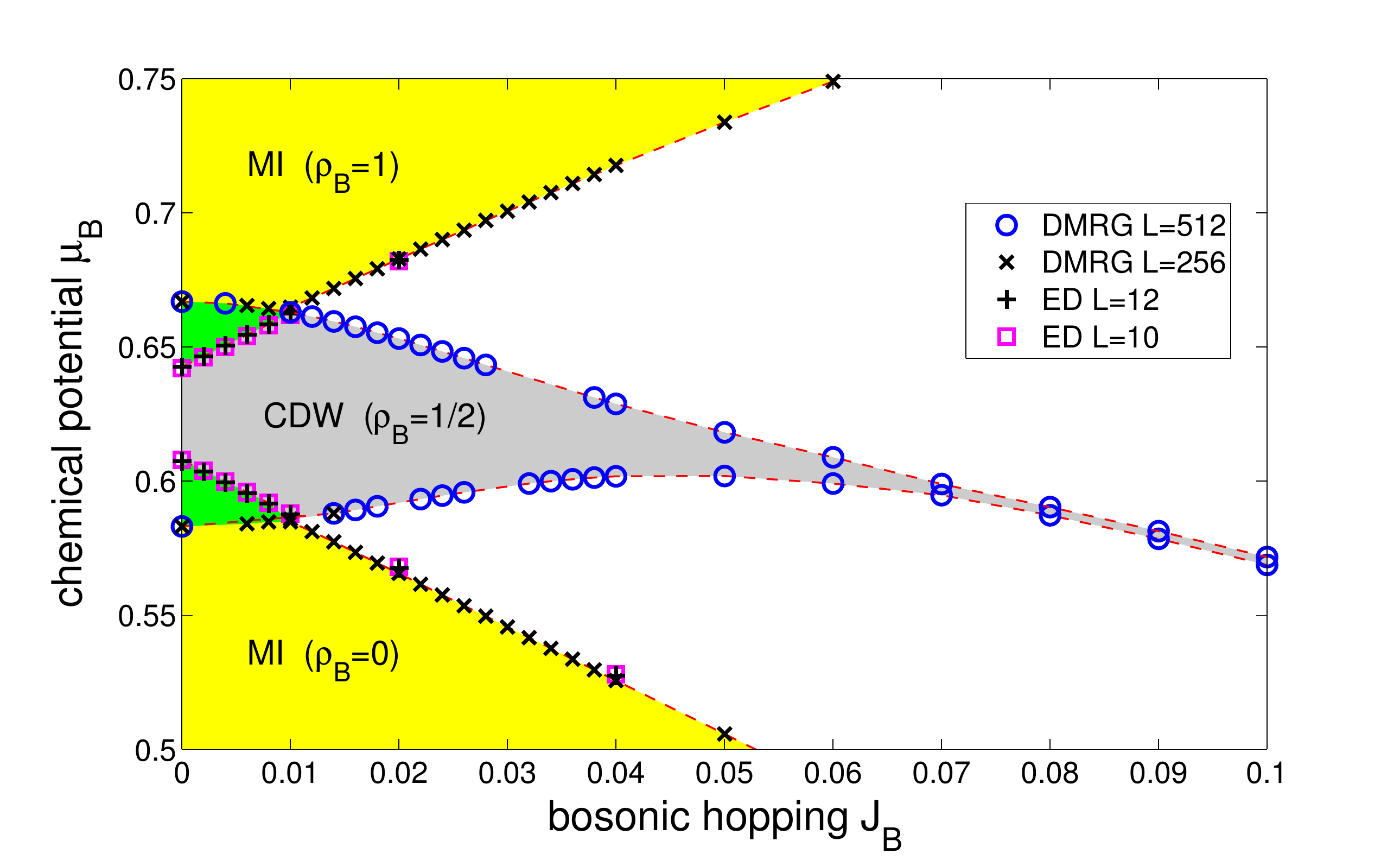}
\caption{%
 Bulk phase diagram of the BFHM for $\rho_F=\frac12$. Beside the expected Mott insulating lobes (yellow), an incompressible CDW for half filling is found (gray). Most prominent feature is the overlap region between the CDW phase and each of the MI (green), indicating a thermodynamic instable region as discussed in the text.  The numerical data (points) were obtained for $V=1.25$ and $J_F=10$, using DMRG and ED for small lattices with system sizes as indicated in the legend. The dashed lines are to guide the eye.
}
\label{fig:PhaseDiagramFastFermions}
\end{figure}

\subsection{limitations of mean-field approximation}

To assess the validity of the fermionic mean-field approximation we calculate the phase diagram for the lowest two lobes by numerical means using DMRG and exact diagonalization (ED) shown in Figure \ref{fig:PhaseDiagramFastFermions}. Different from a simple BHM
the Mott lobes do not touch each other, opening a gap between them.
Within this gap another incompressible phase arises, where the bosonic filling is also one half. This phase can be identified as a charge density
wave at double half filling. The CDW phase extends even beyond the gap between the Mott lobes, partially overlapping with the MI. This overlap region indicates the existence of a thermodynamic unstable phase with coexistence of Mott insulator and CDW. 
Both, the existence and the extent of the CDW and coexistence phases can be fully understood by an effective bosonic theory 
which we will develop in the following sections.

\section{Effective boson model}\label{chap:AdiabaticEliminiation}

\subsection{Adiabatic elimination of the fermions}

In order to understand the phase diagram of Fig.\ref{fig:PhaseDiagramFastFermions} we derive an effective bosonic model. To this end we split the full Hamiltonian \eqref{eq:BFHM} into a bosonic part $\H B$, a fermionic part $\H F$ and an interaction part $\H I$, i.e., $\hat H = \H B + \H F + \H I$, with
\begin{align}
 \H B &=
 -J_B\sum_j\left(\ad_j\a_{j+1}+\ad_{j+1}\a_{j}\right)+\frac{U}{2}\sum_j\nb_j\left(\nb_j-1\right)\label{eq:
BosonicHamiltonian}\\
   \H F &= -J_F\sum_j\left(\hat c^\dagger_j \hat c_{j+1}+ \hat c^\dagger_{j+1} \hat c_{j}\right)+ V 
\sum_j \widetilde n_j \nf_j\label{eq:FermionicHamiltonian}\\
 \H I &= V\sum_j(\nb_j-\widetilde n_j)\nf_j.
\end{align}
At this step, we introduced a bosonic mean-field potential $\widetilde n_j$ in $\H I$. This term will be important later on in the renormalization procedure
discussed in section \ref{chap:RenormalizationFermions} to describe the
backaction of a bosonic CDW onto the fermionic system. For 
the moment, this term is kept without specifying $\widetilde n_j$. The effective bosonic Hamiltonian is obtained by an adiabatic elimination, which is performed in the framework of the scattering matrix 
\begin{equation}
 \hat{\mathcal S} ={ \mathcal T}\exp\left\{-\frac i\hbar \int_{-\infty}^\infty {\rm d}\tau  \H 
I(\tau)\right\} \label{eq:SMatrix}
\end{equation}
of the full system in the interaction picture, i.e., $\H {I}(\tau) = e^{-\frac{i}{\hbar}(\H B +\H F)\tau}\  \H I\  e^{\frac{i}{\hbar} (\H B + \H F)\tau}$ and $\mathcal T$ being the time ordering operator. Tracing out the fermionic degrees of freedom yields the bosonic scattering matrix via $ \hat{\mathcal S} ^{\rm B}_{\rm eff} = \Tr_{\rm F} \hat{\mathcal S}$. 
Neglecting cumulants of the fermionic density higher than second order
in the cumulant expansion $ \left \langle \exp \{ s X \} \right\rangle_{\rm X} = \exp \left\{\sum_{m=1}^\infty \frac{s^m}{m!} 
\langle\langle X^m\rangle\rangle \right\}$ and applying a Markov approximation \cite{Louisell1973,Carmichael1993}, which amounts to 
replacing  the two-time fermion denisty-density correlator with a delta-function in time, we arrive at an effective bosonic interaction Hamiltonian
\begin{eqnarray}
  \H{I}^{\rm eff} &=&   
-J_B\sum_j\left(\ad_j\a_{j+1}+h.a.\right)+\frac{U}{2}\sum_j\nb_j\left(\nb_j-1\right)\notag\\
 &&+ V \sum_j \Bigl(\nb_j-\widetilde n_j\Bigr) \langle\nf_j\rangle_{\rm 
F}\label{eq:effectiveBHMFull}\\
  &&+\sum_{j}\sum_{d=-\infty}^\infty   g_d(\rho_F) \Bigl(\nb_j-\widetilde n_j\Bigr)  
\Bigl(\nb_{j+d}-\widetilde n_{j+d}\Bigr).\notag
\end{eqnarray}
Two different effect of the fermions on the bosonic subsystem become apparent: the fermions induce (i) a mean-field potential (1st order) and (ii) density-density interactions (2nd order). Physically, the second process can be understood as an interaction between bosons mediated by elementary excitations
of the fermionic ground state, which induces a long-range interaction. The corresponding coupling constants at distance $d$ read 
\begin{equation}
 g_d(\rho_F) = - i \frac{V^2}{2 \hbar} \int_{-\infty}^\infty {\rm d}\tau\   \langle\langle\mathcal T\ 
\nf_j(\tau)\nf_{j+d}(0)\rangle\rangle_{\rm F}. \label{eq:CouplingsGeneral}
\end{equation}
Assuming free fermions, i.e. setting $V=0$ in $\H F$,
 the two-time density-density correlation of the fermions can be calculated analytically, which yields
\begin{equation}
 g_d(\rho_F) = -\frac{V^2}{2\pi^2J_F}  \int_ {0}^{\rho_F\pi} \!\!\! \! {\rm d} \xi \int_ {\rho_F \pi}^{\pi} \!\!\! {\rm d} 
\xi^\prime\   \frac {\cos (d\xi)\cos(d\xi^\prime )}{ \cos(\xi)-\cos(\xi^\prime)}. \label{eq:CouplingsFreeFermions}
\end{equation}

Before discussing the phase diagram, several important properties of the coupling constants should be 
mentioned. The first thing to observe is the existence of a particle-hole symmetry for fermions $g_d(\rho_F)=g_d(1-\rho_F)$ which can be seen by substituting $\xi \to \pi-\xi$ and $\xi^\prime \to \pi-\xi^\prime$ and interchanging $\xi \leftrightarrow \xi^\prime$ afterwards. This is a natural consequence of the underlying fermionic system. 

Secondly, for any density $\rho_F\not=0,1$, the local interaction is reduced, i.e.,
\begin{equation}
g_0(\rho_F)\Bigr|_{\rho_F\not=0,1} =  -\frac{V^2}{8J_F}<0.
  \end{equation}
 This negative shift is in full agreement with the results from 
\cite{Buechler2003,Lutchyn2008,Tewari2009}, predicting the enhancement of the superfluid phase because of a reduction of the on-site interaction $U$ of the bosons. Beyond this simple local renormalization, (\ref{eq:effectiveBHMFull})
incorporates further interaction effects modifying the phase diagram.

\begin{figure}[h]%
 \includegraphics*[width=\linewidth]{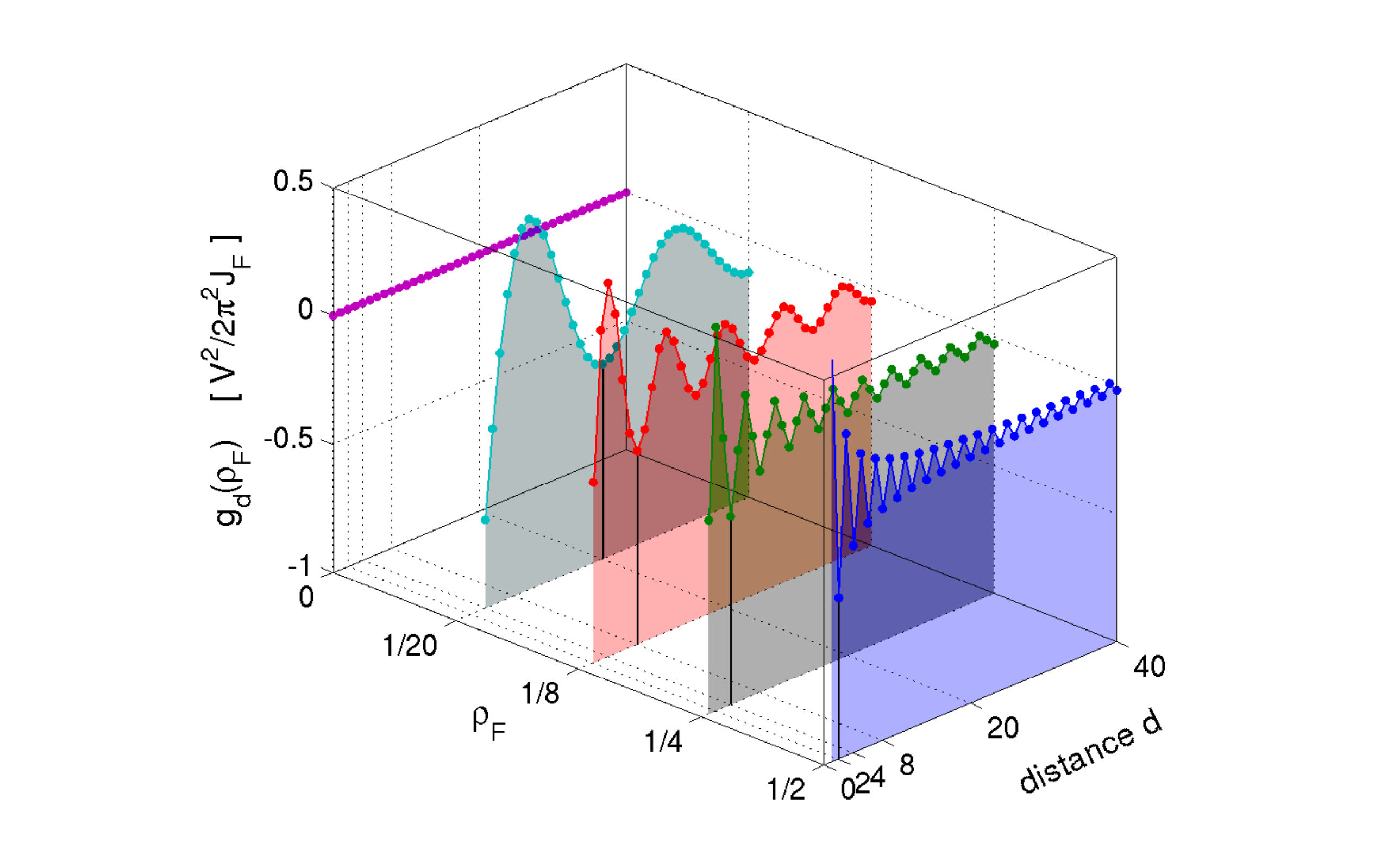}
  \caption[]{%
  Dependence of the coupling strength $ g_d(\rho_F)$ for selected densities 
$\rho_F = 0, 1/20, 1/8, 1/4, 1/2$ on the distance $d$. The periods of the oscillations are $1/\rho_F = \infty, 20, 8, 4, 2$. For all cases, the signs in the minima are negative and the maxima positive with a strict alternation from site to site for the case of half filling. 
}
    \label{fig:CouplingsRealSpaceDistanceFree}
\end{figure}

 Figure \ref{fig:CouplingsRealSpaceDistanceFree} shows the dependence of the couplings 
on the distance $d$ for selected densities $\rho_F$.  One can see a periodic modulation with wavelength $1/\rho_F$ (for $\rho_F<\frac12$, otherwise the wavelength is given by $1/(1-\rho_F)$). This behavior of is typical for induced couplings of the RKKY-type (Rudermann-Kittel-Kasuya-Yosida) \cite{Rudermann1954,Kasuya1956,Yosida1957}. The most interesting case can be found for $\rho_F=1/2$. In this case, the wavelength of $2$ leads to a strict alternation in the sign of the couplings from site to site. As a result, the effective Hamiltonian (\ref{eq:effectiveBHMFull}) displays repulsive nearest-neighbor, attractive next-nearest-neighbor, repulsive next-next-nearest-neighbor interaction and so on. See \cite{Soeyler2009} for a similar, numerical study in this case for two dimensions. Thus the induced long-range coupling provides a simple explanation for the existence 
of a CDW phase  at double half filling $\rho_F=\rho_B=\frac12$ \cite{Pollet2006,Titvinidze2008}.

\begin{figure}[h]%
\includegraphics*[width=\linewidth]{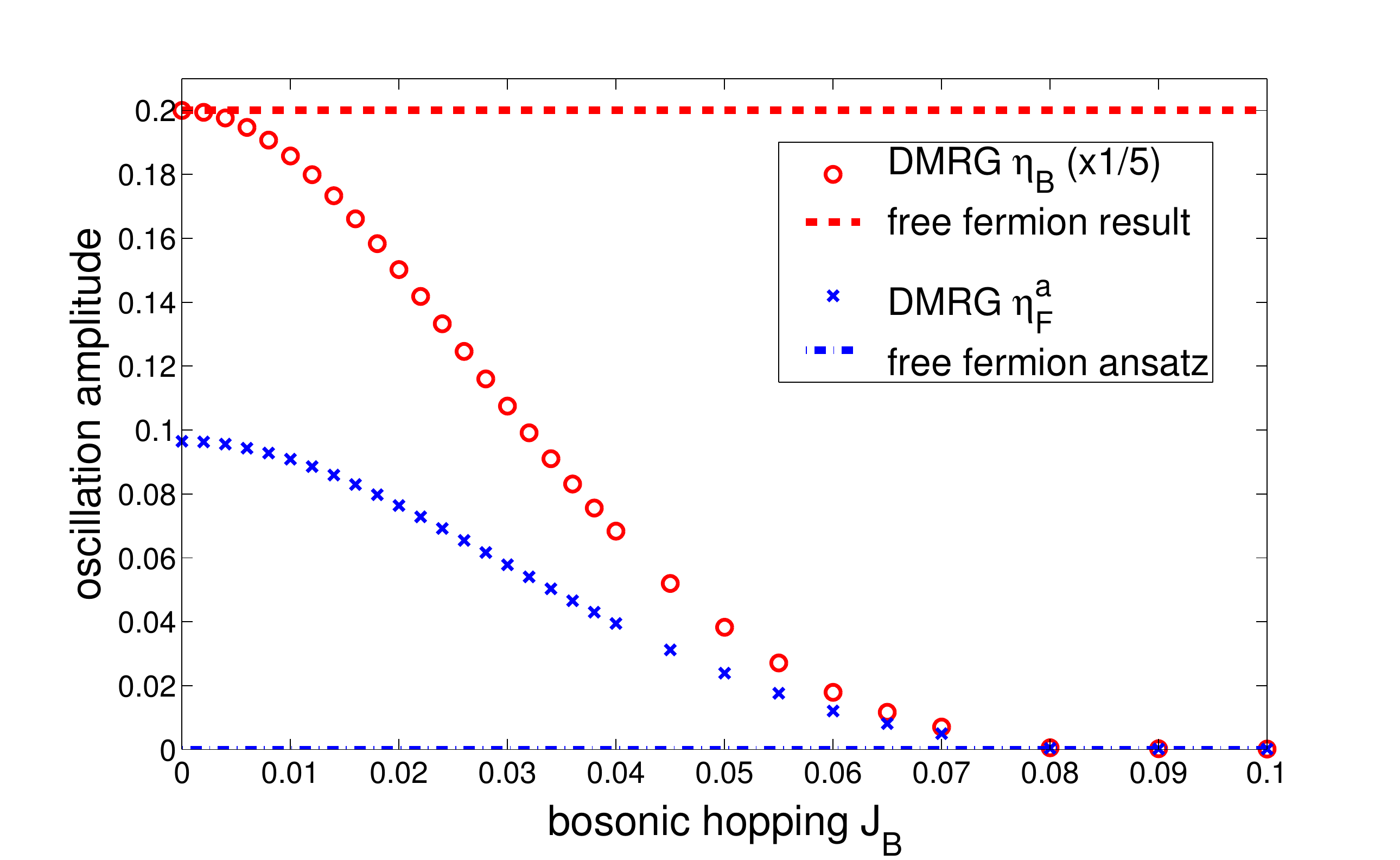}
  \caption[]{%
  Amplitude of the bosonic CDW as a function of the bosonic hopping $J_B$ for 
$V=1.25$ and $J_F=10$. Since the effective theory predicts a CDW for any hopping $J_B$ (dashed lines), the necessity of a renormalization scheme is evident. Additionally the non-zero amplitude of the fermionic CDW is in strong contrast to the underlying ansatz and another indication of a more involved physics. The numerical data are obtained from DMRG for lattice of $512$ sites and $N_F=N_B=256$.
}
    \label{fig:CDWamplitudes}
\end{figure}

A detailed inspection of the coupling constants $g_d$ of the effective model, eq. (\ref{eq:effectiveBHMFull}), reveals some problems, however. 
In particular one finds that the envellope 
of the coupling constants scales inversely with distance $d$,
\begin{equation}
  g_d(\rho_F) \sim \frac{1}{d}.
 \end{equation}
Clearly for very large values of $d$ the effective coupling will be suppressed below this value due to retardation effects ignored when applying the
Markov approximation. But even for moderate values of $d$, where retardation can safely be disregarded this scaling leads to problems.
As mentioned above the existence of a CDW phase results from the oscillatory long-range interactions, which can be seen most easily 
for the case of vanishing bosonic hopping $J_B$.  Adding bosons to 
the system starting from zero filling up to $\rho_B=\frac12$, the first boson occupies an arbitrary site $j$. A second boson minimizes the energy at site $j\pm2$, since here the density-density interaction is negative. All additional particles will continue to occupy even sites, ending up in the CDW phase at half filling $\rho_B=1/2$. However, since the couplings decay as $\frac1d$, the total interaction energy in the thermodynamic limit diverges. The latter argument also holds for $J_B>0$ and thus the ground state would always be a CDW with full amplitude $\eta_B=1$ for {\it any} hopping $J_B$. This result is in strong contrast to the numerical results displayed in Figure \ref{fig:PhaseDiagramFastFermions} and more precisely in Figure \ref{fig:CDWamplitudes}. The latter one shows the amplitude of the bosonic CDW from Figure \ref{fig:PhaseDiagramFastFermions} as a function of the bosonic hopping $J_B$ 
in comparision to the prediction from eq.(\ref{eq:CouplingsFreeFermions}). The Figure also gives a hint to a solution of this problem: Also shown is the amplitude of a fermionic CDW, i.e., the CDW phase is indeed a double CDW. The appearance of fermionic density modulations shows that the initial assumption of free fermions is invalid and the back-action of the bosons needs to be included.
This will be done now in an approximate way by incorporating an oscillatory mean-field $\widetilde n_j$ into the equations of motion of the fermions. The same arguments also hold in the case of a commensurate fermionic density but $\rho_F\not=\frac12$, leading to a ground state which has a boson at every $\frac1{\rho_F}$-th site.

  \section{Renormalization of the fermionic system and effective boson 
Hamiltonian}\label{chap:RenormalizationFermions}\label{chap:RenormalizedHamiltonian}

\subsection{Backaction of slow bosons to fast fermions}

The fermion mediated interaction drives the bosons into a CDW state. This bosonic density wave, in turn, acts as an external potential to the fermionic subsystem, and this backaction leads to a renormalization of the induced boson-boson interaction and thus needs to be taken into account.
 In the following we restrict ourselves to the most interesting case $\rho_F=1/2$, a generalization to other situations with $\rho_F=1/m$ with $m\in\mathbf{N}$ is possible but will not be provided here.
As will be shown the back-action can be modeled to a high accuracy within a simple mean-field
description for the bosons $ \widetilde n_j$ in equation (\ref{eq:FermionicHamiltonian}). \footnote{A similar ansatz is used in \cite{Pazy2005} to study the influence of the wavelength of the bosonic CDW on the fermionic system.}.
\begin{align}
 \widetilde n_j&=\rho_B\left[1+\eta_B (-1)^j\right]\label{eq:AnsatzBosonicCDW}\\
 &=\rho_B(1-\eta_B)+2\rho_B\eta_B\   \delta\bigl( \sin (\frac{\pi j}{2})\bigr),
\end{align}
 Here we introduced the amplitude of the bosonic CDW $\eta_B$ as a free parameter. 
With this mean-field back-action, the fermionic correlators have to be calculated with respect to fermions in an alternating potential, given by
\begin{multline}
 \H F = -J_F\sum_j\left( \hat c^\dagger_j\hat c_{j+1}+\hat c^\dagger_{j+1}\hat 
c_j\right)\label{eq:FermionicHamiltonianRenormalized}\\
 + V2\rho_B\eta_B\sum_j  \delta( \sin \pi \frac j2)\nf_j. 
 \end{multline}
 In this Hamiltonian, an overall energy shift $V\rho_B (1-\eta_B)\rho_F$ is left out.
Resembling free fermions in an oscillatory super-potential
a solution can be found straightforwardly, e.g. by means of a canonical transformation \cite{Rousseau2006,Lieb1961}. The resulting expressions are however rather involved and the quantities needed hard to express. For that reason we employ a Green's function approach, which allows to extract all required quantities for the bosonic Hamiltonian at double half filling in a compact form.

\subsection{Free fermions in an alternating lattice potential}\label{sec:DysonEquation}

The second order cumulant $\langle\langle\nf_j(\tau)\nf_{j+d}(0)\rangle\rangle_{\rm F}$ factorizes by use of Wick's theorem  into a product of advanced and retarded Green's functions ($\tau \ge 0$)
\begin{align}
 \langle\langle\nf_j(\tau)\nf_{j+d}(0)\rangle\rangle_{\rm F} &=\mathcal G^{(+)}_{j,j+d}(t+\tau,t)\   \mathcal 
G^{(-)}_{j,j+d}(t+\tau,t)\notag\\
 &\hspace{-2cm}=  \left\langle \hat c^\dagger_j(t+\tau)\hat c_{j+d}(t)\right\rangle\left \langle \hat 
c_{j}(t+\tau )\hat c^\dagger_{j+d}(t) \right\rangle.\label{eq:DensityDensitySplitted}
\end{align}
The free Green's functions, i.e. in the absence of the boson-induced backaction,  can be obtained by a straightforward calculation, which gives 
\begin{equation}
\mathcal G^{(0\pm)}_{k,k^\prime}(\omega)=\pm\delta_{k,k^\prime}\frac{i}{\sqrt{2\pi}} \frac{1}{\epsilon_k\mp\omega\oplus i\delta} \label{eq:GreensFunctionFreeFermions}
\end{equation}
in the frequency-momentum domain. Here we introduced the dispersion relation $\epsilon_k=-2J_F\cos(2\pi\frac kL)$ of the free particles.
The last term in the denominator is introduced to assure convergence and will be properly removed later on. It is
$\oplus = +$ for $k\in\mathcal K_F$ and $\oplus = -$ for $k\not\in\mathcal K_F$ distinguishing between momentum modes within the Fermi sphere $\mathcal K_F$ and outside.

The full Greens function taking into account the boson-induced potential can be obtained from a simple Dyson equation
\begin{align}
 \mathcal G^{(+)}_{k,k^\prime}(\omega)  &= \mathcal G^{(0+)}_{k,k^\prime}(\omega)\\
 &\hspace{0.1cm}+\frac{i}{\hbar} \sqrt{2\pi}V\eta_B\rho_B\ \mathcal G^{(0+)}_{k,k}(\omega)   
\sum_{\alpha=\pm1}  \mathcal G^{(+)}_{k+\frac L2\alpha,k^\prime}(\omega)\label{eq:DysonGreensAdvanced}\notag
\end{align}
and similarly for $ \mathcal G^{(-)}$ and can be solved analytically.
 Going back to real space and taking the thermodynamic limit gives
\begin{eqnarray}
&&
  \mathcal G^{(\pm)}_{jj+d}(t+\tau,t) = \nonumber\\
  &&\qquad =
   \frac{1}{2\pi}\int_{0}^{\pi}{\rm d}\xi\,\cos( d\xi)\ 
  e^{- i\tilde\epsilon(\xi)\tau} \left(1\pm\frac{\epsilon(\xi)}{\tilde\epsilon(\xi)}\right)\nonumber\\
&&\qquad
-(-1)^j\frac{a}{2\pi}\int_{0}^{\pi}{\rm d}\xi\,\cos(d\xi)\ \frac{e^{- i\tilde\epsilon(\xi)\tau }}{\tilde\epsilon(\xi)}.
\label{eq:GreensFunctionRenormalizedFermions}
\end{eqnarray}
Here we introduced  the normalized fermion energies
\begin{equation}
\epsilon(\xi) = \cos(\xi),\qquad \tilde \epsilon(\xi) = \sqrt{\cos^2(\xi) + a^2}
\end{equation}
and the modulation factor
 \begin{equation}
 a=\frac{V\eta_B\rho_B}{2\hbar J_F}.
 \end{equation}
 Note that the integration cannot be carried out explicitly for arbitrary distance $d$.

The full Green's functions does not only allow to calculate the density-density correlator in equation (\ref{eq:CouplingsGeneral}) but also gives a prediction of the behavior of the fermionic system, as long as the bosonic CDW amplitude $\eta_B$ is known. We first verify the analytic expression of the Green's function in the fermionic problem itself, i.e., all numerical data shown are calculated for the Hamiltonian (\ref{eq:FermionicHamiltonianRenormalized}).

\textbf{Local density:} The expression for the Green's functions gives an (analytic) prediction of the fermionic density in the alternating potential. Using $\langle\nf_j\rangle_{\rm F} =  \mathcal G^{(+)}_{j,j+0}(t+0,t)$, the fermionic density evaluates analytically as
\begin{equation}
  \langle\nf_j\rangle_{\rm F} =\frac{1}{2}- (-1)^j\frac{a}{\pi\sqrt{1+a^2}}\ K\left[\frac{1}{1+a^2} 
\right].
\end{equation}
The first important result from the renormalization procedure therefore is
\begin{equation}
\langle\nf_j\rangle_{\rm F}=\frac12\left[1-\eta_F^a (-1)^j\right],\label{eq:FermionicDensity}
\end{equation}
where $\eta_F^a=\frac{2a}{\pi\sqrt{1+a^2}}K\left[\frac{1}{1+a^2} \right]$ and $K[x]$ is the complete elliptic integral of the first kind \cite{Abramowitz1964}. This means, the renormalization procedure results in the prediction of a fermionic CDW with some amplitude $\eta_F^a$ which is 
entirely determined by the amplitude of the corresponding bosonic CDW $\eta_B$ through the
parameter $a$. The fixed relation between bosonic and fermionic CDW is
in full agreement with the numerical results from Figure \ref{fig:CDWamplitudes}.

Another feature of \eqref{eq:FermionicDensity} which will be important for the later discussion of the full BFHM is the minus sign in front of the site dependent part. This is a direct consequence of the alternating boson potential ansatz. Since the interaction $V$ is chosen positive, i.e., repulsion between bosons and fermions, it is expected that the phase of the bosonic and fermionic density wave is shifted by $\pi$ compared to each other. For the case of attractive interaction, both density waves are in phase. This is in full agreement with the numerical results. In the limit $a\to 0$, corresponding to the free fermion case the result for the density reduces to the result for free fermions at half filling, i.e., $ \langle\nf_j\rangle_{\rm F}=\frac12$.

  \begin{figure}[h]%
\includegraphics*[width=\linewidth]{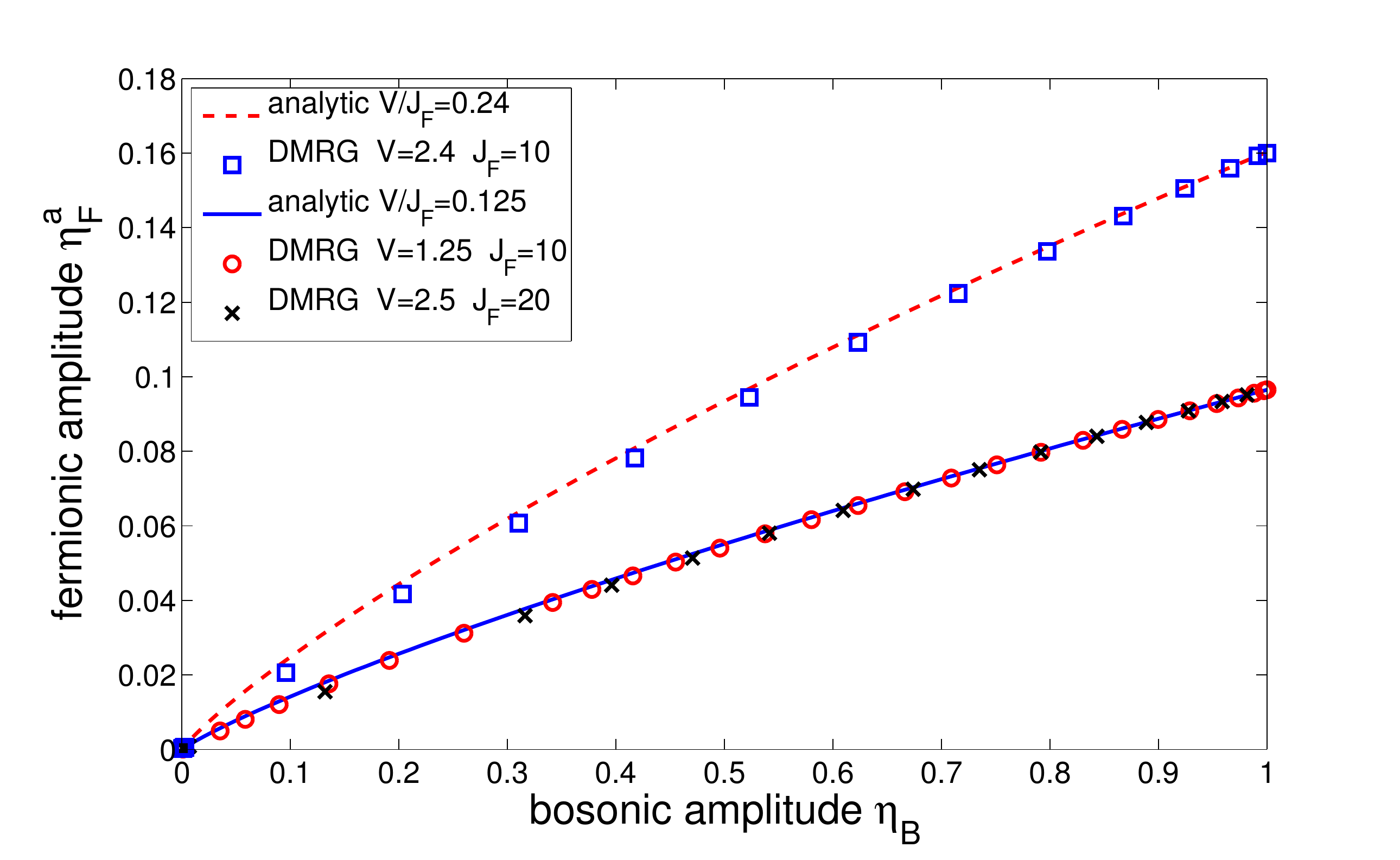}
  \caption[]{%
  Amplitude $\eta_F^a$ of the fermionic CDW versus the bosonic amplitude 
$\eta_B$ for different numerical data. Shown are the numerical results (data points) presented in Figure \ref{fig:CDWamplitudes} and for $V=2.4$ and $J_F=10$ obtained from the full BFHM. The solid lines are the analytic results for $\eta_F^a$.
  }
    \label{fig:CDWamplitudes2}
\end{figure}

  \begin{figure}[h]%
\includegraphics*[width=\linewidth]{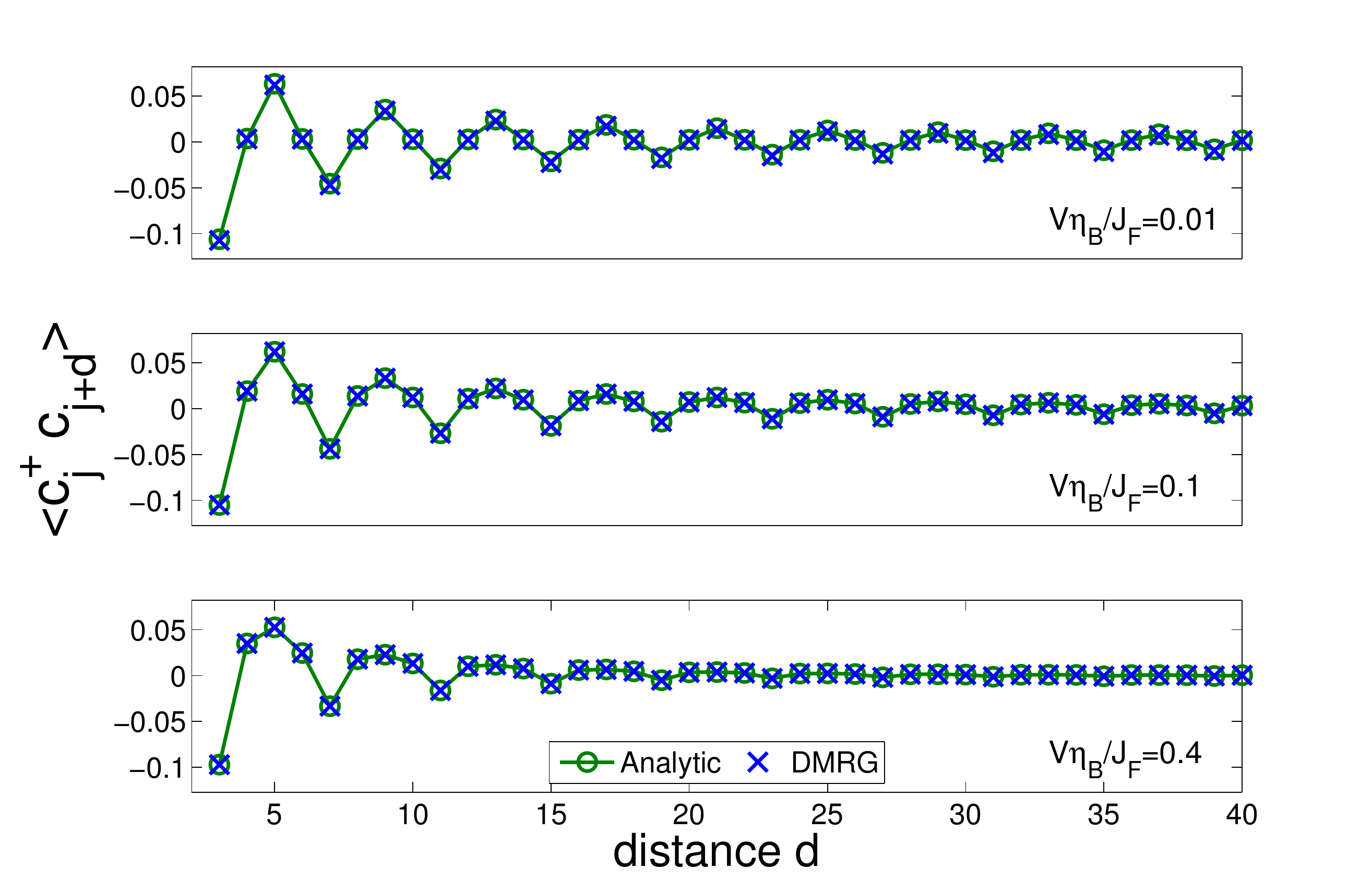}
  \caption[]{%
 Distance dependence of the first-order correlations $ <\hat c^\dagger_j\ \hat 
c_{j+d}> $  for three different interactions $V\eta_B$ calculated from the fermion model (\ref{eq:FermionicHamiltonianRenormalized}). Solid lines are the theoretical results from a numerical integration of \eqref{eq:GreensFunctionRenormalizedFermions}. }
    \label{fig:FermionicCorrelationsa}
\end{figure}

\

  \textbf{First-order correlations:} Figure \ref{fig:FermionicCorrelationsa}  shows numerical results for the  first-order correlations $  \bigl\langle\hat c^\dagger_j\ \hat c_{j+d}\bigr\rangle =   \mathcal G^{(+)}_{j,j+d}(t+0,t)$ compared to the analytic results. Unfortunately, the integral expression for the Green's function cannot be evaluated analytically for arbitrary distance $d$, making a numerical integration necessary.  The perfect agreement proves the validity of the solution (\ref{eq:GreensFunctionRenormalizedFermions}).
    
  \
    
  \textbf{Density-density correlations:}  Finally we calculate the density-density correlations used in 
the expression for the coupling constants (\ref{eq:CouplingsGeneral}) with the renormalized fermionic model. Having a closer look at the result for the Green's function (\ref{eq:GreensFunctionRenormalizedFermions}) it can be seen that they are of the general form $\mathcal G^{(\pm)}_{j,j+d}(t+\tau,t) = A_\pm - a B$. Since the density cumulant split up into products of advanced and retarded Green's function they can thus be written as
  \begin{equation}
    \langle\langle\nf_j(\tau)\nf_{j+d}(0)\rangle\rangle_{\rm F} = A_+ A_- - a(A_++A_-) + a^2 
B^2.\label{eq:SimplifikationDensityDensity}
  \end{equation}
 From the definition of the coupling constants (\ref{eq:CouplingsGeneral}) we can see, that they are 
proportional to $V^2$. This means, that in order $V^2$, only the first term in \eqref{eq:SimplifikationDensityDensity} is relevant.

Following these argument, the renormalized form of the density-density cumulant reads
  \begin{align}
  \langle\langle \hat m_j(t+\tau) \hat m_{j+d}(t)\rangle\rangle 
&=\frac{1}{4\pi^2}\int_{0}^{\pi}\int_{0}^{\pi}{\rm d}\xi{\rm d}\xi^\prime\,\notag\\
 & \hspace{-1.5cm} \times\cos(d\xi )\,\cos(d\xi^\prime ) e^{-i\bigl(\tilde\epsilon(\xi)+\tilde\epsilon(\xi^\prime)\bigr)\tau}
\label{eq:CumulantRenormalizedFermions}\\
 &\hspace{-1.5 cm}\times\left(1+ \frac{\epsilon(\xi)}{\tilde\epsilon(\xi)}\right)
 \left(1- \frac{\epsilon(\xi^\prime)}{\tilde\epsilon(\xi^\prime)}\right) .\notag
   \end{align}
  This is the main result from the renormalization procedure. Comparing this result to that 
of free fermions (at $\rho_F=\frac12$) one can see, that the corresponding limit $a\to0$ gives the same result. Note, that the last line in (\ref{eq:CumulantRenormalizedFermions}) serves as a cutoff function which constrains the  integration limits to the free fermion values in the limit $a\to0$.

\subsection{Renormalized Hamiltonian}

\begin{figure}[htb]%
  \includegraphics*[width=\linewidth]{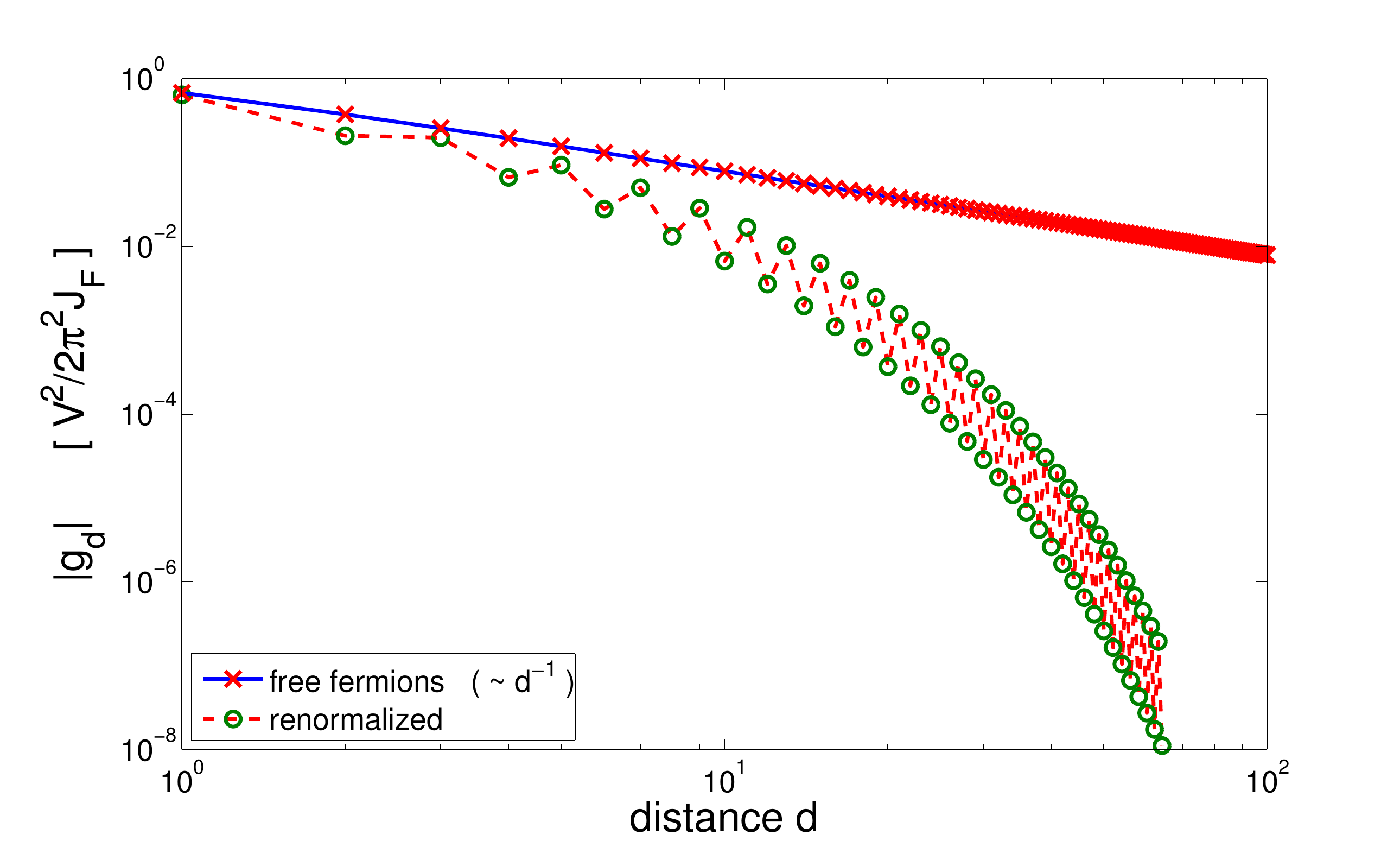}%
  \caption[]{%
 Comparison of the couplings for the free fermion case ($a=0$) and the 
renormalized couplings for $a=0.1$. The free fermion couplings decay as $\frac 1d$, whereas the renormalized couplings decay much faster, preventing the divergence of the energy for the ground state.}
      \label{fig:ComparisonCouplings}
\end{figure}

Applying the time integration from \eqref{eq:CouplingsGeneral}  the renormalized couplings  $g_d(a)$ at half fermionic filling $\rho_F=1/2$ can be found to be
\begin{eqnarray}
  g_d(a) &=& -\frac{V^2}{4 \pi^2}\int_{0}^{\pi}\int_{0}^{\pi}{\rm d}\xi{\rm d}\xi^\prime\,  
\frac{\cos(d\xi)\,\cos(d\xi^\prime)}{\tilde\epsilon(\xi) +\tilde\epsilon(\xi^\prime)}\nonumber\\
&& \times \left(1+ \frac{\epsilon(\xi)}{\tilde\epsilon(\xi)}\right)
 \left(1- \frac{\epsilon(\xi^\prime)}{\tilde\epsilon(\xi^\prime)}\right).
\label{eq:CouplingsRenormalized}
\end{eqnarray}
Since we restricted ourselves to the case of half filling for the fermions, the additional argument $\rho_F$ is dropped here but the dependence of the renormalized couplings on the amplitude factor $a$ is explicitly  written.  Figure \ref{fig:ComparisonCouplings} shows a comparison of the couplings from the free free fermion case to the case $a\sim\eta_B>0$. Obviously the decay is much faster than $1/d$ thus resolving the divergence. 

The knowledge of the renormalized couplings finally allows to write down the effective bosonic Hamiltonian for half fermionic filling $\rho_F=1/2$. Starting from \eqref{eq:effectiveBHMFull} together with the renormalized fermionic density \eqref{eq:FermionicDensity}, the couplings \eqref{eq:CouplingsRenormalized} and the ansatz for the bosonic CDW \eqref{eq:AnsatzBosonicCDW}, the full effective bosonic Hamiltonian is given by
 \begin{multline}
  \H B^{\rm eff} =
-J_B\sum_j\left(\ad_j\a_{j+1}+\ad_{j+1}\a_{j}\right)+\frac{U}{2}\sum_j\nb_j\left(\nb_j-1\right)\\
  -\bar\mu\sum_j \nb_j-\Delta\sum_j\nb_j (-1)^j +\sum_j\sum_d g_d(a)\  
\nb_j\nb_{j+d}\label{eq:effectiveHamiltonian}.
 \end{multline}
 Beside the usual hopping and interaction terms, two prominent features arise. On the one hand, the 
already discussed long-range density-density interaction with couplings  $g_d(a)$ lead to the emergence of CDW phases. These are further stabilized by the induced alternating potential with amplitude
  \begin{equation}
  \Delta=2\rho_B\eta_B \widetilde g_\pi(a)+V \eta_F^a/2,\label{eq:InducedAlternatingPotential}
 \end{equation}
 being a direct consequence of the fermionic density wave
\begin{equation}
 \langle\nf_j\rangle_{\rm F}=\frac12\left[1-\eta_F^a (-1)^j\right].
\end{equation}
Although derived only for the case of double-half filling, the emergence of the induced chemical
potential
 \begin{equation}
  \bar\mu=2\rho_B \widetilde g_0 (a)-V/2 \label{eq:InducedChemicalPotential}
 \end{equation}
in combination with the general ansatz also allows for an extension of the effective Hamiltonian to other fillings $\rho_B$. The  amplitude factor $a=\frac{V\eta_B\rho_B}{2\hbar J_F}$ 
i.e. the amplitude $\eta_B$ of the induced bosonic CDW is still a free parameter. For the Fourier transform, the identities
\begin{equation}
 \sum_d g_d(a) = \widetilde g_0(a),\hspace{1cm} \sum_d (-1)^d\ g_d(a) = \widetilde g_\pi(a),
\end{equation}
hold. For the two momenta $k=0$ ad $k=\pm \pi$ analytic expressions for the Fourier-transformed couplings can be found. %
\begin{align}
    \widetilde g_{\pm\pi}(a)  &= -\frac{V^2}{8\pi J_F}\frac{1}{\sqrt{1+ a^2}} 
\left(2K\Bigl[\frac{1}{1+a^2}\Bigr]-E\Bigl[\frac{1}{1+ a^2}\Bigr]\right)\\
    \widetilde g_0(a)  &=-\frac{V^2}{8\pi J_F }\frac{1}{\sqrt{1+ a^2}} E\Bigl[\frac{1}{1+a^2}\Bigr],
  \end{align}
with $E[x]$ being the complete elliptic integral of second kind \cite{Abramowitz1964}.

Before we exploit the resulting Hamiltonian in the determination of the phase diagram, possible approaches in a self-consistent determination of the bosonic CDW amplitude are 
discussed in the next section.

\subsection{Self-consistent determination of $\eta_B$} \label{chap:SelfconsistentAmplitude}

The introduction of the bosonic CDW amplitude $\eta_B$, or respectively the amplitude factor $a$ as a free parameter demands a proper 
procedure to fix its value. Although the knowledge of $\eta_B$ as a function of the bosonic hopping $J_B$ is not necessary in the discussion of the phase diagram as done in our approach, a possible reproduction of Figure \ref{fig:CDWamplitudes} would further support the validity of our approach. To this end
we will discuss different variational ansatz functions for the ground state and determine the
CDW amplitude by minimizing the energy.

\

\begin{figure*}[htb]%
 \includegraphics*[width=.49\textwidth]{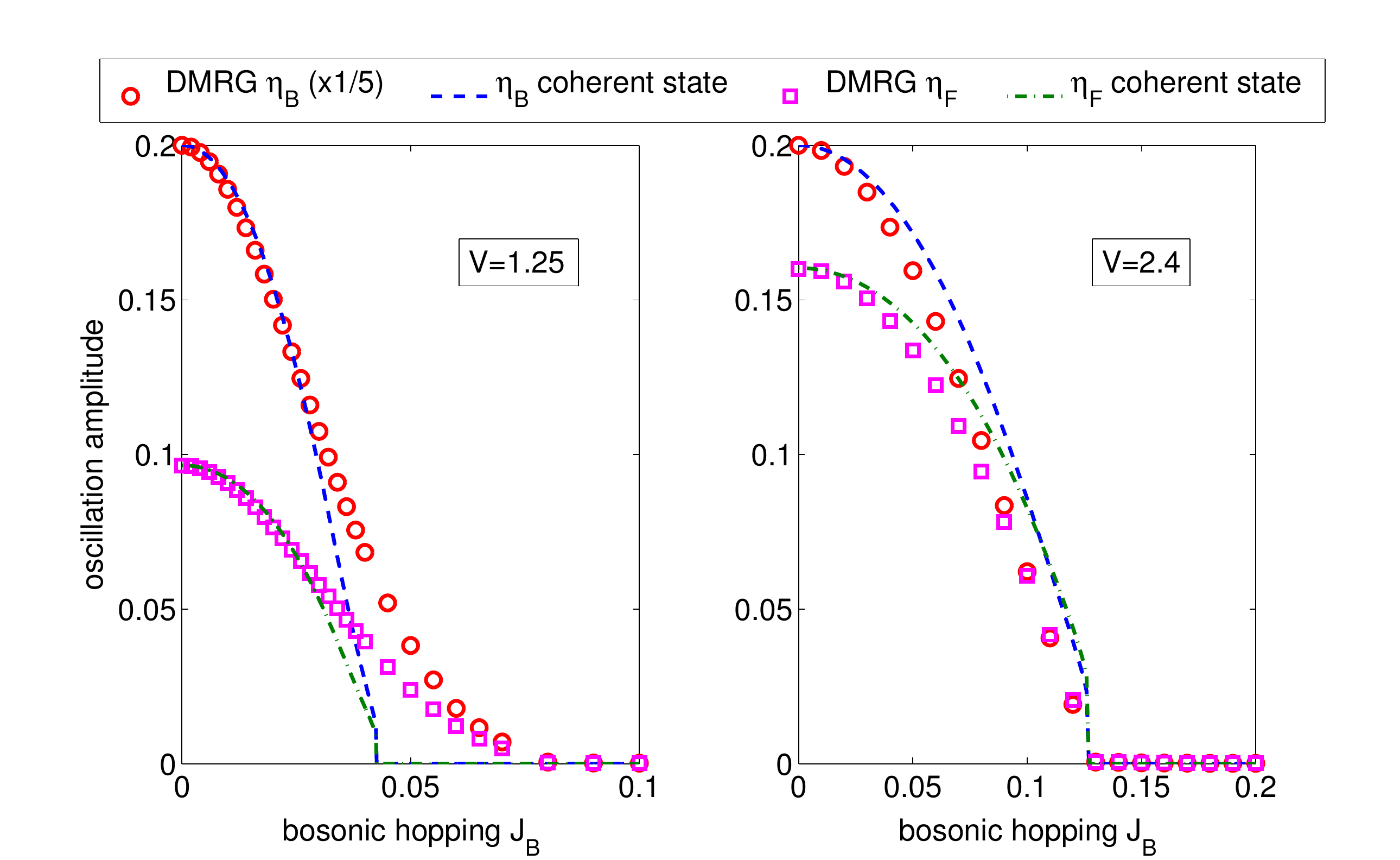}
\includegraphics*[width=.49\textwidth]{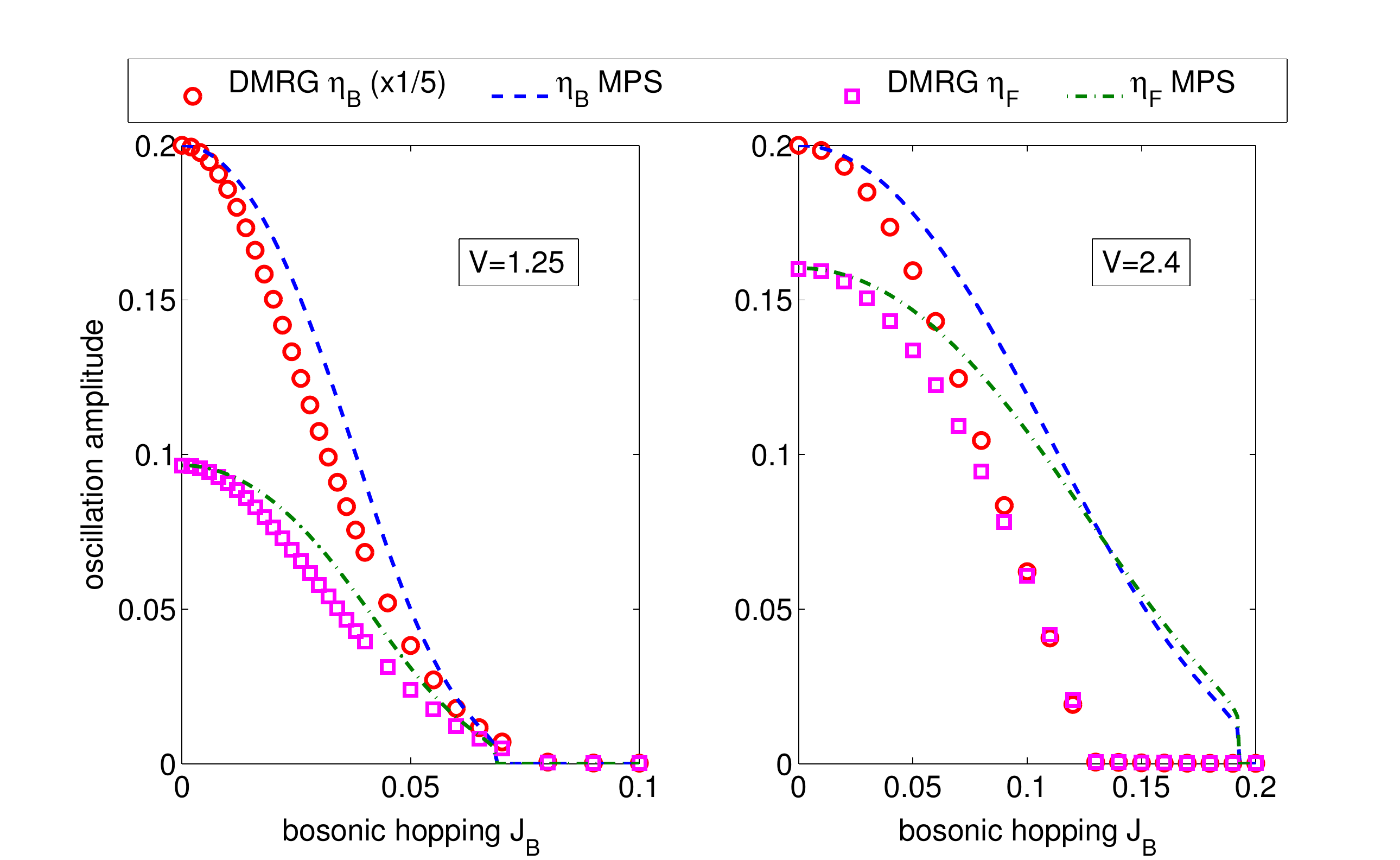}
  \caption[]{%
 (left) Self-consistent determination of the amplitude of the bosonic CDW from the 
minimization of the energy for the effective Hamiltonian with respect to a coherent state ansatz. Shown are the same numerical results as in Figure \ref{fig:CDWamplitudes} (left plot, $L=512$) as well as results for $V=2.4$ and $J_F=10$ (right, $L=256$). One can see the rather poor quantitative agreement with a general qualitative agreement.
(right) Self-consistent determination of the amplitude of the bosonic CDW from the 
minimization of the energy for the effective Hamiltonian with respect to a matrix product state ansatz. Shown are the same numerical results as for Figure \ref{fig:CDWamplitudea} (left plot, $L=512$) and results for $V=2.4$ and $J_F=10$ (right, $L=256$). One can see the better quantitative agreement compared to the result for the coherent state in Figure \ref{fig:SelfconsistentCoherentAmplitude} for small amplitude factor $a$, i.e., for small interaction $V$.
 }
    \label{fig:SelfconsistentAmplitude}
\end{figure*}

\textbf{Coherent state:}
The simplest choice for the ground state of Hamiltonian \eqref{eq:effectiveHamiltonian} is given by local coherent states $\left|\alpha\right\rangle$ with alternating
amplitude: 
\begin{equation}
\left| \Psi \right\rangle^{\rm coh} = \prod_{j=-\infty}^\infty   \left|\alpha_+\right\rangle_{2j}\left|\alpha_-\right\rangle_{2j+1}.
\end{equation}
With this the local densities read
\begin{equation}
 ^{\rm coh}\left\langle \Psi\right| \nb_j \left|\Psi\right\rangle^{\rm coh} = \frac12\left[1+\eta_B 
(-1)^j\right]
\end{equation}
and $\alpha_\pm=\sqrt{\frac12\pm \frac12\eta_B}$. The variatioal energy 
$E\left[\eta_B\right]=\ ^{\rm coh} \left\langle\Psi\right| \H B^{\rm eff} \left|\Psi\right
\rangle^{\rm coh}$ now becomes a function of $\eta_B$ and upon neglecting unphysical contributions from the interaction \footnote{Although the coherent state incorporates all Fock states $n$, for the treated CDW only states with $n\le 1$ are of importance.}, the energy is given by
\begin{align}
E\left[\eta_B\right] &= -J_B \sqrt{1-\eta_B^2} +\frac12 g_0(a)\\
&\hspace{1cm}- \frac14V\eta_F^a \eta_B-\frac14 \widetilde g_a(\pi) \eta_B^2-\frac14 \widetilde g_a(0).\notag
\end{align}
We stress that the amplitude factor $a=\frac{V\eta_B}{4\hbar J_F}$ as well as the fermionic amplitude $\eta_F^a$ also depend on $\eta_B$. Minimization of this function with respect to $\eta_B$ at the end gives a prediction of the bosonic CDW amplitude. This is shown in Figure \ref{fig:SelfconsistentAmplitude}, where the self-consistent prediction is compared to the numerical data from Figure \ref{fig:CDWamplitudes} and to data for $V=2.4$ and $J_F=10$. One can see that the coherent approach gives a qualitatively good agreement for small $J_B$, but the quantitative agreement is rather poor in particular for larger $J_B$ because of the strongly simplified ansatz used here.

\

\textbf{Matrix product state:}
Better results for the CDW amplitude may be found from a minimal matrix product like ansatz. 
\begin{equation}
\left| \Psi \right\rangle^{\rm MPS} = \prod_{j=-\infty}^\infty   \sum_{i_1,i_2=0}^1 A_{i_1i_2}\ \left|i_1\right\rangle_{2j}\left|i_2\right\rangle_{2j+1},
\end{equation}
which also eliminates problems arising from the higher number states. With the prefactors $A_{i_1i_2}$ which are chosen to be real, we introduce four free parameters which have to be minimized in general.  This set of parameters can be reduced by constraints from the normalization of the ground state as well as the expected local densities \eqref{eq:AnsatzBosonicCDW}. Altogether, these constraint reduce to $A_{00}=A_{11}\equiv0$ and the energy functional only depends on $\eta_B$ as
   \begin{equation}
    \begin{split}
      E[\eta_B] &=  - J_B \sqrt{1-\eta_B^2} -\frac V2\eta_F^a \eta_B-\frac12 \widetilde g_a(\pi) 
\eta_B^2-\frac12 \widetilde g_a(0)\\
  &\hspace{1cm}+ (1-\eta_B^2)\bigl[\frac12 g_0(a)-\frac12 g_1(a)\bigr].
    \end{split}
   \end{equation}
 The corresponding numerical results for the minimization can be found in Figure 
\ref{fig:SelfconsistentAmplitude}. The quantitative agreement is slightly better compared to the coherent state approach for smaller interaction $V$ but still the strong simplification of the ansatz pays its tribute. For larger $V$, the matrix product ansatz seems to fail. Nevertheless, the two 
procedures to self-consistently determine the amplitude $\eta_B$ show that this free parameter can in principle be calculated with more sophisticated ansatzes. 
 
\section{Phase diagram of the effective boson model}\label{chap:ResultsFastFermionsFullModel}

We now use the effective bosonic Hamiltonian 
 \begin{multline}
  \H B^{\rm eff} =
-J_B\sum_j\left(\ad_j\a_{j+1}+\ad_{j+1}\a_{j}\right)+\frac{U}{2}\sum_j\nb_j\left(\nb_j-1\right)\\
  -\bar\mu\sum_j \nb_j-\Delta\sum_j\nb_j (-1)^j +\sum_j\sum_d g_d(a)\  
\nb_j\nb_{j+d}\label{eq:effectiveHamiltonian2}
 \end{multline}
to calculate the full phase diagram and compare it to the numerical results from Figure \ref{fig:PhaseDiagramFastFermions}. As a reminder, the potentials $\bar\mu$ and $\Delta$ are given by 
 \begin{equation}
  \bar\mu=2\rho_B\widetilde g_0(a)-V/2, \hspace{0.5 cm}  \Delta=2\rho_B\eta_B \widetilde g_\pi(a)+V 
\eta_F^a/2.
 \end{equation}
 The calculation of the phase boundaries of the different incompressible regions (MI, CDW) is performed
by determining the particle-hole gap for fixed particle number. For an incompressible phase with filling $\rho_B$, the chemical potentials of the upper and lower boundaries are obtained from
 \begin{equation}
  \mu_{\rho_B}^\pm = \pm \Bigl[E(\rho_BL\pm1)-E(\rho_BL) \Bigr].
 \end{equation}
First we restrict ourselves to the zero-hopping limit $J_B=0$. Later on we employ degenerate perturbation theory in $J_B$. It should be mentioned, that both, in the zero hopping limit as well as in the small hopping region to a very good approximation $\eta_B=1$. 

\subsection{Zero-hopping phase diagram}

\begin{figure*}[htb]%
 \includegraphics*[width=.49\textwidth]{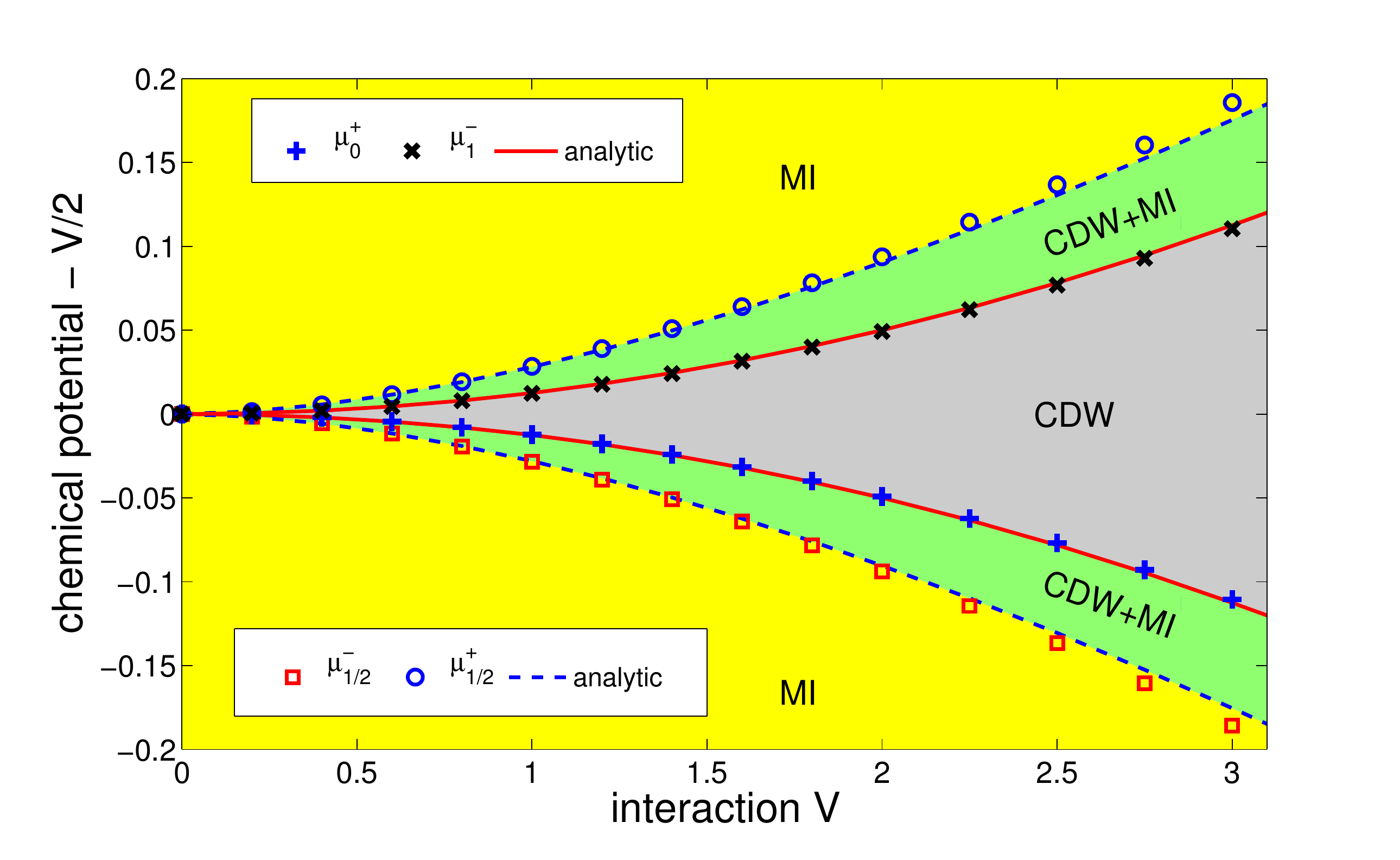}
\includegraphics*[width=.49\textwidth]{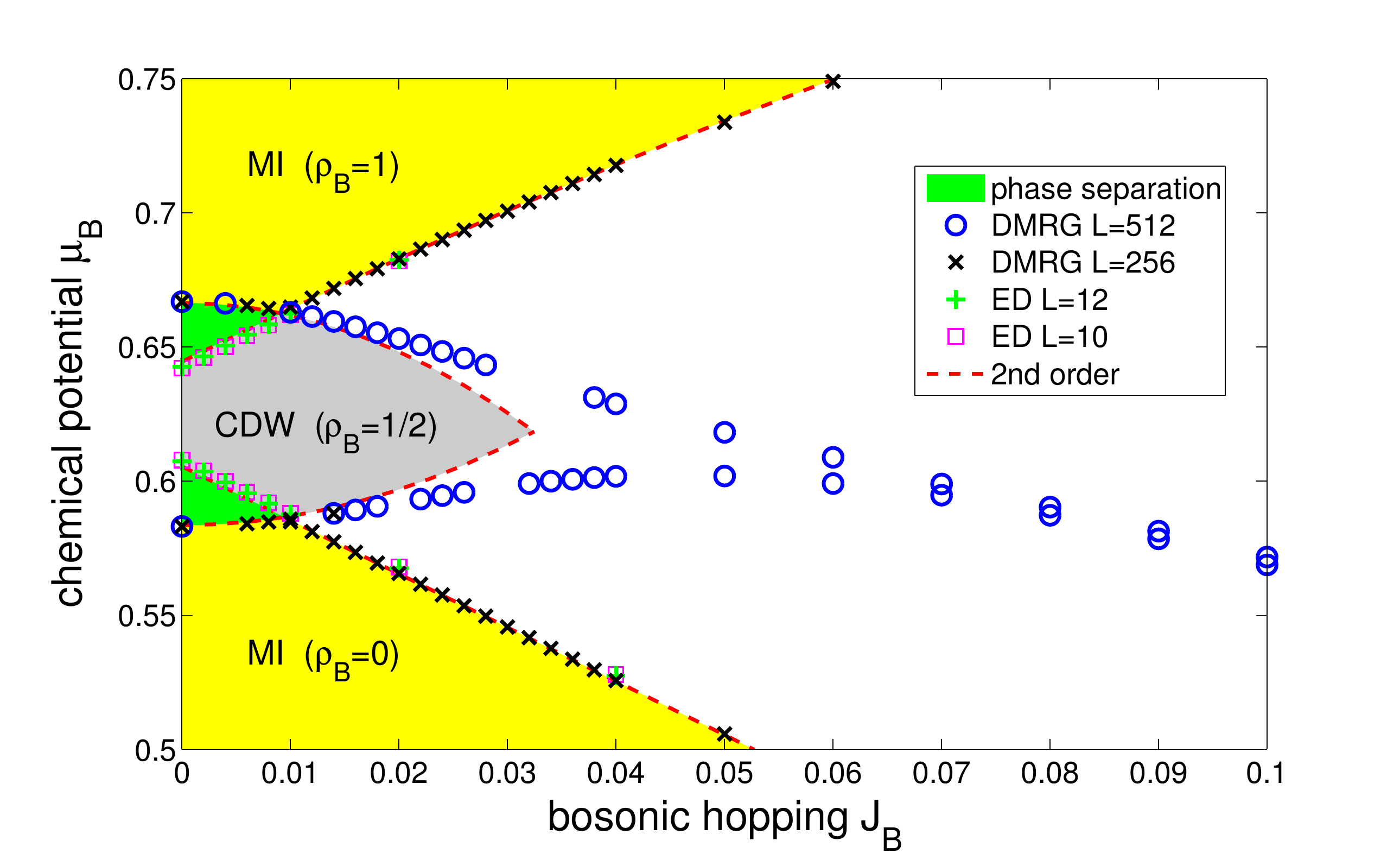}
  \caption[]{%
 (left) Phase diagram of the effective bosonic Hamiltonian for vanishing bosonic 
hopping $J_B=0$. Data points are the numerical results obtained from DMRG and ED for the full BFHM and the lines are the analytic results. Yellow: extended region of the different Mott insulators (MI). Gray: charge density wave phase (CDW); Green: Coexistence region of CDW and MI. Fermionic density $\rho_F=1/2$ and $J_F=0$.
(right) 
Analytic results for the phase diagram together with the numerical results 
from Figure  \ref{fig:PhaseDiagramFastFermions}. The agreement between the analytics and the numerics is quite reasonable with the natural deterioration for larger hopping $J_B$ due to the perturbative treatment.
 }
    \label{fig:AnalyticPhaseDiagram}
\end{figure*}

The calculation of the chemical potentials for $J_B=0$ is straightforward. In this case, the energy is given by a replacement of the number operators $\hat n_j$ in \eqref{eq:effectiveHamiltonian2} by real numbers according to the ground state in the system. Additionally, the density $\rho_B$ and a possible CDW amplitude $\eta_B$ needs to be fixed. This is done for the Mott insulator with unity filling $(\rho_B;\eta_B;\left\langle\hat n_j\right\rangle)=(1;0;1)$, the empty Mott insulator $(0;0;0$) as well as the CDW $(\frac12;1;\frac12[1+(-1)^j])$ and the corresponding particle or hole states. 
This gives
 \begin{align}
 \mu^-_1 &=\frac V2 - g_0(0),\\
  \mu^\pm_{\frac12} &= \frac V2\pm\frac V2\eta_F^a \pm g_0(a),\\
   \mu^+_0 &= \frac V2 +  g_0(0)  , 
 \end{align}
 which together with the results for the couplings $g_d(a)$ \eqref{eq:CouplingsRenormalized} and the 
fermionic CDW amplitude $\eta_F^a$  from \eqref{eq:FermionicDensity} allow to construct the phase diagram at vanishing bosonic hopping. This is shown in Figure \ref{fig:AnalyticPhaseDiagram}, where the chemical potentials are displayed as a function of the interaction $V$ for a fixed fermionic hopping $J_F$, where the mean-field shift $\frac V2$ is subtracted. 
 
One recognizes from Figure \ref{fig:AnalyticPhaseDiagram}
 a very good agreement between the numerical results of the full BFHM and the analytic results 
obtained from the effective bosonic Hamiltonian. Increasing deviations for larger $V$ could be addressed both to the breakdown of the Markov approximation as well as the negligence of higher order contributions in \eqref{eq:CumulantRenormalizedFermions}. Most prominent feature is the overlap between the MI and CDW phases, i.e., $\mu_0^+> \mu_{\frac12}^-$ and $\mu_1^-< \mu_{\frac12}^+$. This behavior, already seen in Figure \ref{fig:PhaseDiagramFastFermions},  indicates a \emph{negative compressibility}
 \begin{equation}
 \kappa = \frac{\partial \langle\hat N\rangle}{\partial \mu_B} < 0
 \end{equation}
within the coexistence phase. Coexisting phases are not new (see e.g. \cite{Batrouni2000,Titvinidze2008,Hubener2009,Soeyler2009}), but the coexistence of a Mott insulator and a CDW phase has to our knowledge not been reported before. A physical explanation of this effect can easily be given. In the grand-canonical ensemble, the phase does not exist since in this situation, the number of particles is chosen such that the energy is minimized: which drives the system always into a CDW phase within this region. From a canonical point of view, adding further particles to the CDW phase results in configurations, where the repulsive contribution to the energy remains constant whereas the attractive one is increased; the energy per particle is thus reduced.

\subsection{2nd order strong-coupling expansion}

Going beyond the zero-hopping limit, we perform a perturbation expansion in the hopping amplitude $J_B$. This allows to generate the full phase diagram in the $(\mu_B,J_B)$ plane. Since the methodology of the perturbation theory is quite involved, we only present the basic ideas.  Different formulations of degenerate perturbation theory exist (e.g., as \cite{Freericks1996}  used in  for the pure and disordered Bose Hubbard model), where we use Kato's expansion \cite{Klein1974,Teichmann2009,Eckardt2009}, which relies on the calculation of an effective Hamiltonian (in arbitrary order) within the degenerate subspace. 
Up to second order, Kato's expansion is given by
\begin{equation}
 \hat H^{\rm eff} = E_0+\mathcal P \hat H_1 \mathcal P + \mathcal P \hat H_1\mathcal Q\frac{1}{E_0-\hat 
H_0}\mathcal Q \hat H_1\mathcal P,\label{eq:Kato}
\end{equation}
where $\mathcal P$ is the projector onto the degenerate subspace, $\mathcal Q=\mathbf{1}-\mathcal P$ the orthogonal projector and $E_0$ is the zero order energy of the manifold. Here, the Hamiltonian is written in the form $ \hat H = \hat H_0 + \hat H_1$, where $\hat H_1$ is the perturbation, i.e., the hopping in our case. For the calculation of the effective Hamiltonian, only the action of \eqref{eq:Kato} on any input state $\ket \Psi_l$ from the degenerate subspace needs to be studied. In our case, the result is of the form
\begin{align}
 \hat H^{\rm eff} \ket\Psi_l &= E_0(\ket\Psi_l) + J_1\Bigl[ \ket\Psi_{l-1}+\ket\Psi_{l+1}\Bigr]\\
 &\hspace{1cm}+ J_2 \Bigl[ \ket\Psi_{l-2}+\ket\Psi_{l+2}\Bigr]+W \ket\Psi_l\notag
\end{align}
since our perturbation only consists of a nearest-neighbor hopping. A generalization to arbitrary long-range hopping can be done. This (maximally) tridiagonal matrix representation of the effective Hamiltonian can be solved by a Fourier transform, which gives the energy
\begin{equation}
 E = E_0 + 2 J_1 \cos(2\pi\frac{k}{L}) + 2 J_2 \cos(4\pi\frac{k}{L}) +W,
\end{equation}
where the $k$ mode has to be chosen such that the energy is minimal. In this system this is typically the case for $k=0$ since both $J_1\sim J_B$ and $J_2\sim\frac{ J_B^2}{E_0-\langle\hat H_0\rangle}$ are negative. A crucial point in the calculation comes from the nature of the effective bosonic Hamiltonian in \eqref{eq:effectiveHamiltonian2}. Since the density-density interaction is long ranged, the energy denominator depends on the distance of the particle performing the first hopping process from the reference site where the additional particle (hole) is situated. This needs to be taken into account for the calculation of the chemical potentials.

A major difficulty is the dependence of the results on coupling strengths $g_d(a)$ up to a large distance $d$. For the analytic results used in Figure  \ref{fig:AnalyticPhaseDiagram} it turns out, that $d\approx100$ is sufficient to gain convergence. Here we only give the numerical values for the chemical potential. Directly plugging in numbers, these are given by
\begin{align}
 \mu_0^+ &=0.605469-2 J_B,\\
 \mu_{\frac12}^-&=0.583612+33.076 J_B^2,\\
 \mu_{\frac12}^+&=0.666388- 45.4392 J_B^2,\\
 \mu_1^-&=0.644531+2 J_B-4.12927 J_B^2
\end{align}
Figure \ref{fig:AnalyticPhaseDiagram} shows the previously used numerical data from Figure \ref{fig:PhaseDiagramFastFermions} together with the analytic predictions. The overall agreement to a second order treatment is quite reasonable. Altogether, our analytic approach allows to completely derive the bosonic phase diagram analytically and provides an intuitive physical understanding.

\subsection{effects of open boundaries}

In the above discussion we have considered infinite systems or systems with periodic boundary conditions. The situation becomes more interesting if effects of confinement are taken into account, which will be discussed in the following.

\begin{figure}[t]
 \includegraphics*[width=\linewidth]{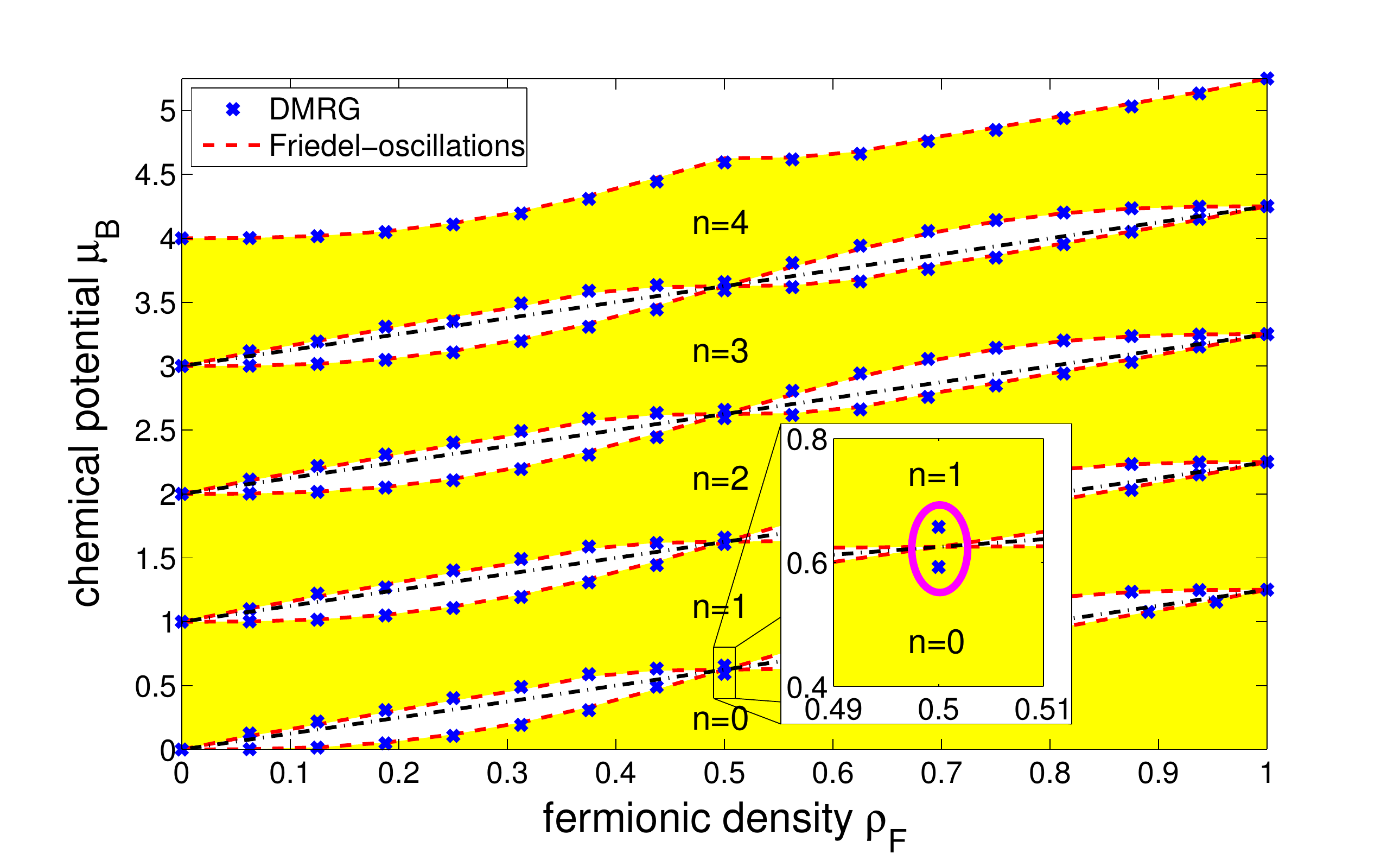}
   \caption{Phase diagram of the BFHM for zero bosonic hopping $J_B=0$ as a function of 
the fermionic filling 
$\rho_F$ at large hopping $J_F=10$. The yellow shaded regions represent the different Mott lobes.  The numerical data are 
obtained for $L=64$ with $V=1.25$ and the agreement with the analytic prediction from the Friedel oscillations (dashed line along the data) 
is very good. The mean-field shift $V\rho_F$ is indicated by the straight dash-dotted lines.  The zoom indicates that contrary to the
fermionic mean-field predictions the Mott lobes do not touch 
for half fermionic filling as discussed previously in the main text.}
   \label{fig:ZeroHoppingFriedelOscillation}
    \end{figure}

In the presence of a confinement, most prominently for open boundary conditions, already the mean-field ground state of the fermions is changed in a very 
important way. Here, the fermionic density displays Friedel oscillations \cite{Friedel1952}, given by 
\begin{equation}
 \left\langle \nf_j\right\rangle =  \frac{N+\frac12}{L+1}-\frac{1}{2(L+1)}\frac{\sin\left(2\pi 
j\frac{N+\frac12}{L+1}\right)}{\sin\left(\frac{\pi j}{L+1}\right)}.\label{eq:FriedelOscillation}
\end{equation}
Thus, instead of a resulting homogeneous chemical potential $\mu_B$ for the bosons, the bosons experience a site-dependent potential $\sum_j \mu_j \nb_j$, with 
$\mu_j =\mu_B- V   \left\langle \nf_j \right\rangle$. This introduces a qualitatively new feature to the system which is equivalent to 
the disordered Bose-Hubbard model (dBHM), or respectively a superpotential BHM. 
Due to the superpotenial the phase diagram in the limit  $J_B=0$ is modified, as can be seen in Figure \ref{fig:ZeroHoppingFriedelOscillation}.
In particular the MI regions do not touch each other anymore in contrast to the BHM with shifted chemical potential.
Considering particle-hole excitations \cite{Freericks1996}, we find for the upper and lower critical chemical potentials for the $n-$th Mott insulator
  \begin{align}
   \mu_n^+ &= V n \rho_F + V \min_j \left\langle \nf_j \right\rangle,\\
    \mu_n^- &= V n \rho_F + V \max_j \left\langle \nf_j \right\rangle.
  \end{align}
  
  \begin{figure*}[t]%
 \includegraphics*[width=.49\textwidth]{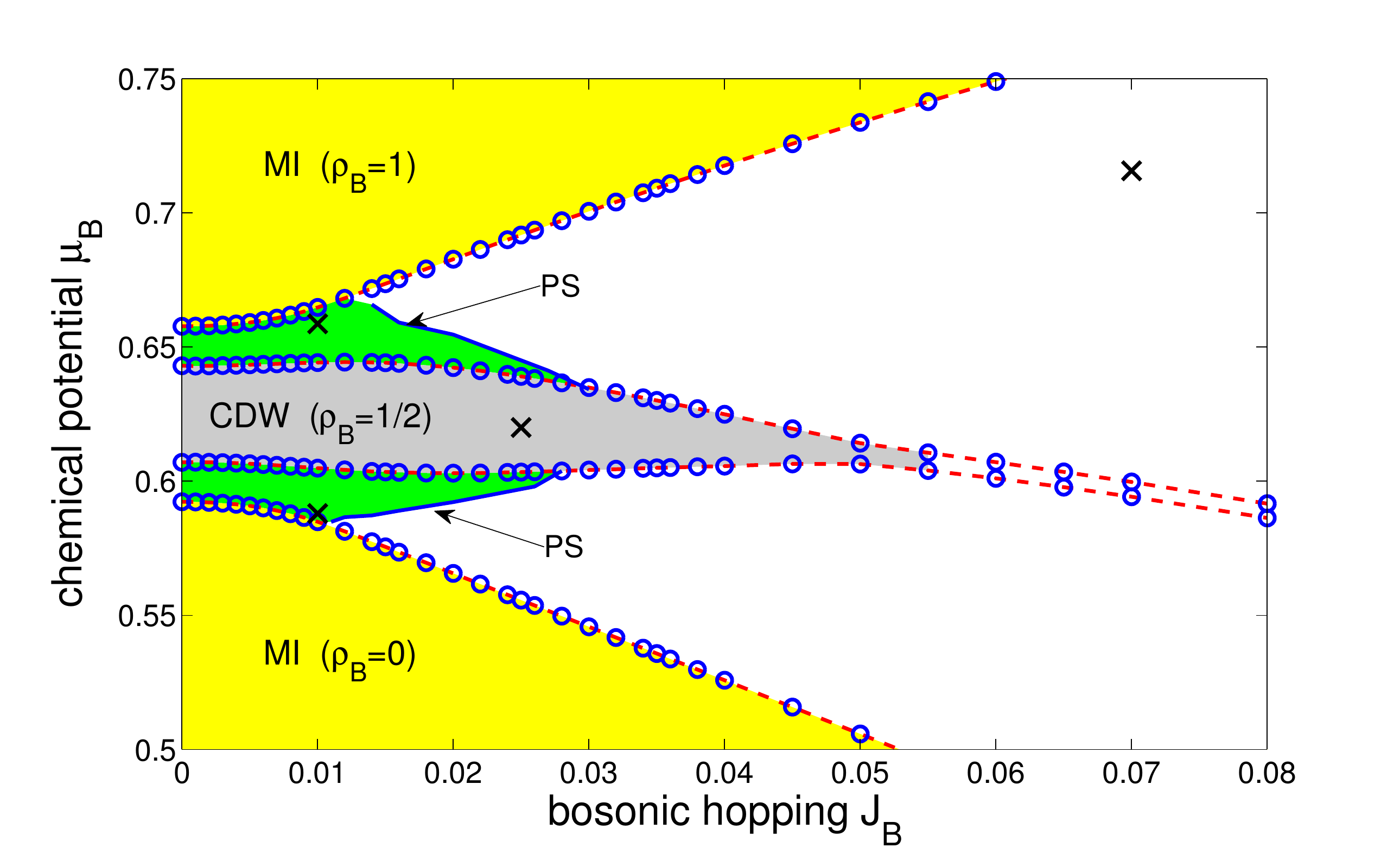}
\includegraphics*[width=.45\textwidth]{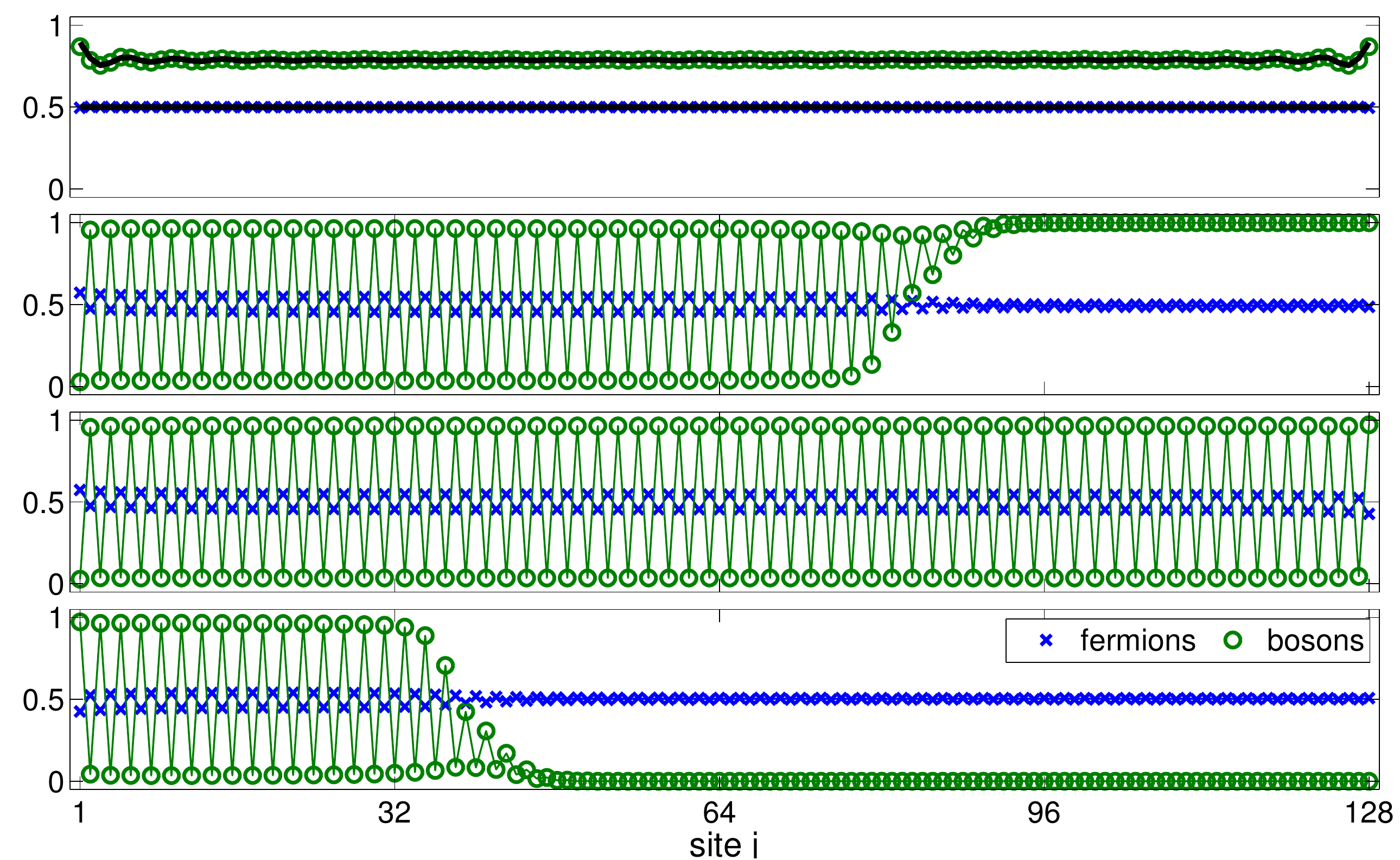}
  \caption[]{%
 (left) Phase diagram of the full BFHM with open boundaries. One can see that the lobes 
bend apart from each other. 
The data are obtained with DMRG and open boundary conditions for a fixed length of $L=128$ sites. The other parameters are $J_F=10$ and $V=1.25$.  
The dashed lines are to guide the eye.
(right) Density profile obtained by DMRG for various numbers of particles. From 
bottom to top: $N_B=20,64,86,101$ for a system of $L=128$ sites. The lower three are for $J_B=0.01$ and the uppermost for $J_B=0.07$. One can immediately see the pinning of the additional particles to the boundary resulting in a phase separation of Mott insulator and CDW. In the uppermost plot, the fermionic state is roughly given by a homogeneous distribution according to the Friedel oscillations whereas the bosons behave as interacting bosons. This can be seen from the additional solid line which, gives the density profile for the same choice of parameters but without interspecies interaction, decoupling the bosons and the fermions. The positions of the data set for the density cuts in the phase diagram are depicted by the small marks in the left figure.
 }
    \label{fig:PhaseDiagramFiniteSize}
\end{figure*}

The phase diagram of the bosonic subsystem is shown in Figure \ref{fig:PhaseDiagramFiniteSize} for $L=128$ and open boundaries .
 The long-range character of the fermion mediated interactions leads to a substantial modification of the dynamics even for relatively large systems. This can directly be seen for the case of the CDW phase, where we first discuss the zero hopping case. Adding a further particle to the CDW phase, this particle has to choose an odd side. Due to the open boundaries the translational symmetry is broken. Thus it matters whether the particle is added close to the boundary or at the center. Because of the long-range interaction, the possible choices differ in energy. The additional energy close to the boundary is given by $\sum_{d=0}^{L/2} g_{2d+1}$, in contrast to the energy at the center $ \sum_{d=-L/4}^{L/4} g_{2d+1}$. The energy is minimal for a position close to the boundary. Adding further particles, the same arguments apply and increasing the filling, a Mott insulating region is growing from the boundary. Switching to small, but finite hopping does not change the situations. As long as the hopping is small compared to the energy difference between the state with a particle pinned close to the border and the state with the additional particle at the center, the reduction of the interaction energy due to pinning 
to the boundary dominates the increase of the kinetic energy. When removing a particle from the system, i.e., going below half filling, the same arguments apply.

This behavior supports our observation of a \emph{phase separation between a Mott insulator and a CDW} 
in the infinite system with negative compressibility. However, the (open) boundary leads to a different dependence of the compressibility, now being strict positive $\kappa  > 0$. This can be seen from Figure \ref{fig:PhaseDiagramFiniteSize}, where the DMRG  results for a system exposed to open boundaries are shown. In contrast to Figure \ref{fig:PhaseDiagramFastFermions}, the Mott lobes and the CDW phase bend apart from each other, not overlapping anymore. This is due to the positive compressibility due to boundaries. The positive compressibility could also be seen in Figure \ref{fig:PhaseDiagramFiniteSize}, where the bosonic filling is shown for three different cuts at fixed $J_B$ along the $\mu_B$-axis. 
The filling is in each situation a monotonous function of $\mu_B$, i.e. $\kappa > 0$. The incompressible CDW and MI phases are clearly observable. Interestingly our system dose not display a so-called {\it Devil's staircase} as described in \cite{Capogrosso-Sansone2010,Bak1982} for the case of a dipolar Bose gas with density-density interactions decaying as $g_d\sim\frac{1}{d^3}$. Most likely, this is because of the alternating sign in our coupling constants together with the alternating potential, where a detailed discussion of this fact might be an interesting supplement to the present work.

\section{Conclusion and outlook}

Deriving an effective bosonic Hamiltonian we provided a comprehensive understanding of the bosonic phase diagram of the Bose-Fermi-Hubbard model in the limit of ultrafast fermions. For double half filling, the physics is dominated by fermion-induced long-range density-density interactions alternating in sign, leading to the emergence of a bosonic charge-density wave phase. 
A naive calculation of induced coupling assuming free fermions leads to divergencies
which are overcome by a renormalization scheme that includes the back-action of the bosonic CDW on the fermions. The effective theory allows for a calculation of the CDW amplitude in very good agreement
with numerical DMRG simulations of the full BFHM.
Beyond half filling, the induced interactions lead to thermodynamically unstable regions in the $(\mu_B,J_B)$-phase diagram, i.e. a phase separation between CDW and Mott insulator. 
Application of the effective theory to Bose-Bose of Fermi-Fermi mixtures is straightforward.


\subsection*{acknowledgement}
This work has been supported by the DFG through the SFB-TR 49, project number 31867626. 
We also acknowledge the computational support from the NIC at FZ J{\"u}lich and thank 
U. Schollw\"ock for his DMRG code. Furthermore we thank B. Capogrosso-Sansone, S. Das Sarma, E. Altmann, W. Hofstetter, C. Kollath, M. Snoek and T. Giamarchi for useful discussions.


\bibliography{BibTex_Alexander_Mering}

\end{document}